\documentclass[%
  aps,
  pre,
  reprint,
  amsmath,amssymb,
  superscriptaddress
]{revtex4-2}

\usepackage{graphicx}
\usepackage{dcolumn}
\usepackage{bm}
\usepackage{hyperref}
\usepackage[caption=false]{subfig}
\usepackage{verbatim}

\usepackage{xcolor,ulem}

\begin{document}

\title{
 A Fokker–Planck approach to a stochastic multiplicative wealth model \\with taxation and  redistribution 
}
\author{Iago Nascimento Barros}
\affiliation{Departamento de F\'isica, Universidade Federal de Vi\c{c}osa (UFV), Vi\c{c}osa, MG, Brazil}
\affiliation{Ibitipoca Institute of Physics (IbitiPhys), Concei\c{c}ão do Ibitipoca, MG, Brazil}

\author{Marcelo Lobato Martins}
\affiliation{Departamento de F\'isica, Universidade Federal de Vi\c{c}osa (UFV), Vi\c{c}osa, MG, Brazil}
\affiliation{Ibitipoca Institute of Physics (IbitiPhys), Concei\c{c}ão do Ibitipoca, MG, Brazil}

\author{Celia Anteneodo}
\affiliation{Departamento de F\'isica, Pontif\'icia Universidade Cat\'olica do Rio de Janeiro (PUC--Rio), Rio de Janeiro, RJ, Brazil}

\date{\today}

\begin{abstract}

We develop a Fokker–Planck description of the dynamics of wealth distribution in a stochastic multiplicative economic growth model with taxation and redistribution, as introduced by P.M.C. de Oliveira.  Extending the original formulation, our theoretical framework includes general redistribution protocols, encompassing a broad class of state-dependent transfer mechanisms. As a particular case, we investigate a two-state protocol designed to emulate conditional cash transfer programs. 
Analytical expressions for the stationary wealth distributions are derived, revealing how the interplay between multiplicative noise, taxation, and redistribution shapes the emergence of inequality. The theoretical results are corroborated by agent-based simulations.
To quantify and compare the impact of the different protocols, we employ the Gini index as a measure of inequality. Our analysis highlights how specific nonuniform redistribution schemes can significantly mitigate wealth disparities.

\end{abstract}

\maketitle

\section{Introduction}

Stochastic multiplicative economic growth is a minimal and commonly used mechanism to account for the emergence of broad (heavy-tailed) wealth distributions~\cite{Bouchaud2000,deOliveira2017} . Its appeal lies in the coexistence of analytical tractability and direct numerical implementation, enabling distribution-level predictions from stochastic differential equations and their associated Fokker–Planck equations, as well as agent-based simulations. A key issue in this line of research is about the existence (and characterization) of a stationary wealth distribution. Indeed, multiplicative accumulation can readily generate heavy tails and high inequality, yet whether such distributions represent genuine steady states is a separate question. Achieving stationarity typically requires additional mechanisms that counterbalance multiplicative amplification, thereby stabilizing the distribution in the long-time limit.

A common way to counteract unchecked multiplicative economic growth is to introduce taxation and redistribution. Together, these mechanisms effectively generate mean-reverting forces (typically through a drift toward the mean) and can stabilize the dynamics, often yielding stationary states that display Pareto-like upper tails. A paradigmatic example is the model proposed by de Oliveira, in which multiplicative growth is followed by a wealth-dependent taxation and a redistribution protocol ~\cite{deOliveira2017,deOliveira2020}. More generally, redistribution plays a central role in ensuring stationarity across a wide range of econophysics models, including wealth-exchange (yard-sale–type) dynamics~\cite{Lima2022,Polk2021} and other agent-based frameworks~\cite{Calvelli2023}, highlighting its robustness as an inequality-mitigating mechanism. In contrast, in the absence of taxation or redistributive policies, such economic dynamics tend to amplify inequality, frequently leading to ever-widening distributions

Many previous studies have considered the particular case of uniform redistribution schemes; see, for example, Refs. \cite{Boghosian2017,deOliveira2017,deOliveira2020,Bouchaud2000,Toscani2008,Iglesias2020, Lima2022}. In realistic settings, however, redistribution is rarely uniform: policies may favor the poor (progressive) or, conversely, advantage the rich (regressive). Moreover, recent works suggest that the functional form of the upper tail near criticality is not universal, but instead depends sensitively on the specific details of public policies \cite{Lima2022,Polk2021}. Understanding how nonuniform redistribution shapes the stationary distribution and the transition toward extreme inequality is therefore of direct interest.

Here we introduce a class of nonuniform redistribution schemes into the P.~M.~C.~de~Oliveira model~\cite{deOliveira2017}, described in Section~\ref{sec:model}, and develop a Fokker--Planck description for the rescaled wealth \(y=W/\langle W\rangle\) at the condensation threshold (where most of the population tends to accumulate in the poorest wealth layer). 
For specific redistribution rules, exact stationary solutions are obtained, and the corresponding theoretical predictions are corroborated by agent-based simulations.
Our results further show that important stylized features of stationary wealth distributions remain robust across a broad class of redistribution mechanisms.

\section{Model}
\label{sec:model}

Following P.M.C. de Oliveira~\cite{deOliveira2017}, we consider a population of $N$ agents, where each agent \(i\) has a wealth \(W_{i,t}\) at time \(t\). At each discrete time step the wealth evolves according to a three-step process:  first, it is multiplied by an idiosyncratic stochastic growth factor; second, a fraction of the resulting wealth is taxed and pooled with the collected revenue, and third, the pool is  redistributed among the agents. This three-step protocol leads to the mapping
\begin{equation}
W_{i,t+1}=\Bigl[1-\tau_{i,t}\Bigr] f_{i,t} W_{i,t}
+ r_{i,t}\sum_{j=1}^N \tau_{j,t} f_{j,t} W_{j,t},
\label{eq:mapping}
\end{equation}
for each agent $i$, where \(f_{i,t}\) is the stochastic return of the wealth of agent $i$ at time $t$, modeled as
\begin{equation}
f_{i,t}=\nu+\varepsilon \eta_{i,t},
\end{equation}
where \(\nu>0\) is the mean return, \(\varepsilon>0\) sets the noise amplitude, and \(\eta_{i,t}\) are identically and independently distributed standard Gaussian variables; \(\tau_{i,t}\) is the tax rate, and \(r_{i,t}\) specifies the fraction of total tax revenue assigned to agent \(i\), which is the mechanism of wealth redistribution.

In the original proposal~\cite{deOliveira2017}, the tax rate was taken to be linear in the wealth share, that is,
\begin{equation}
\tau_{i,t}=A+p\,\omega_{i,t}, \qquad
\omega_{i,t}=\frac{W_{i,t}}{\sum_{k=1}^N W_{k,t}},
\end{equation}
with \(A\in(0,1)\), \(p\in(-1,1)\),  and \(0 \le A + p \le 1\) to ensure \(0\le \tau_{i,t}\le 1\). 
In fact, the parameter \(p\) plays a role analogous to an effective temperature, controlling the transition between a condensed phase (\(p\le0\)) and a non-condensed phase (\(p>0\)) in the absence of redistribution ($r_{i,t} = 0$) \cite{deOliveira2017}. In turn, $r_{i,t}$ plays the role of an external field conjugated to the order parameter that destroys the frozen-active transition.  

In this work, we focus on the critical point \(p=0\) (uniform tax rate), in which case Eq.~\eqref{eq:mapping} reduces to
\begin{equation}
W_{i,t+1}
= (1-A)\,f_{i,t}W_{i,t}
+ r_{i,t}\,A\sum_{j=1}^N f_{j,t}W_{j,t}.
\label{eq:mapping_p0}
\end{equation}
This choice removes the wealth-dependent (nonuniform) component of the tax of the original proposal, thereby isolating the impact of the redistribution rule \(r_{i,t}\). 

We investigate how different redistribution schemes reshape the dynamics defined by Eq.~\eqref{eq:mapping_p0} using analytical methods, and compare the resulting predictions with the results from 
 agent-based simulations.
These simulations are performed by 
iterating the (discrete-time) $N$-dimensional 
stochastic map defined by Eq.~(\ref{eq:mapping_p0}), starting from an egalitarian initial condition with $W_{i,0}=10$ for all $i$. At each iteration, time is incremented by one unit.

Unless differently stated, we use \(N=10^4\) agents, average growth rate $\nu=1.05$, 
noise amplitude $\varepsilon =0.1$, and Gaussian shocks \(\eta_{i,t}\). Stationary distributions are sampled after a transient and averaged over \(M=10^2\) independent realizations of the stochastic map.

\section{Dynamics without redistribution}

In the absence of redistribution (\(r_{i,t}\equiv 0\)), the dynamics for each agent decouples and the general mapping \eqref{eq:mapping_p0} reduces to
\begin{equation}
W_{t+1}=(1-A)\,\bigl(\nu+\varepsilon\eta_t\bigr)\,W_t,
\label{eq:mapping_no_redist_p0}
\end{equation}
with \(\eta_t\sim\mathcal{N}(0,1)\). Defining \(X_t=\ln W_t\) and expanding for \(\nu\gg\varepsilon\), we arrive at
\begin{equation}
\Delta X_t \equiv X_{t+1}-X_t \approx \ln\!\bigl[(1-A)\nu\bigr]+\frac{\varepsilon}{\nu}\eta_t-\frac{1}{2}\left(\frac{\varepsilon}{\nu}\right)^2\eta_t^2.
\end{equation}
Averaging over the noise yields an effective drift
\begin{equation}
C \equiv \langle \Delta X_t\rangle
=\ln\!\bigl[(1-A)\nu\bigr]-D,
\end{equation}
where we define
\begin{equation}
D \equiv \frac{1}{2}\left(\frac{\varepsilon}{\nu}\right)^2,
\end{equation}  
and \(\mathrm{Var}[\Delta X_t]\approx 
2D\). 
Taking the continuous-time limit in the It\^o convention \cite{Gardiner2009}, we obtain
\begin{equation}
\mathrm{d}X_t=C\,\mathrm{d}t+
\sqrt{2D}\,\mathrm{d}B_t,
\label{eq:sde_X_no_redist_p0}
\end{equation}
where \(B_t\) is a standard Wiener process with unit variance. The associated Fokker--Planck equation for the probability density function (PDF) $P_X(X,t)$ reads
\begin{equation}
\partial_t P_X = -\partial_X\!\left(C\,P_X\right) + D\,\partial_X^2 P_X.
\label{eq:fp_X_no_redist_p0}
\end{equation}
For the homogeneous initial condition (\(W_{i,0} = W_0\)), the solution is Gaussian at all times.

\begin{figure}[b!]
    \centering
    \includegraphics[width=\columnwidth]{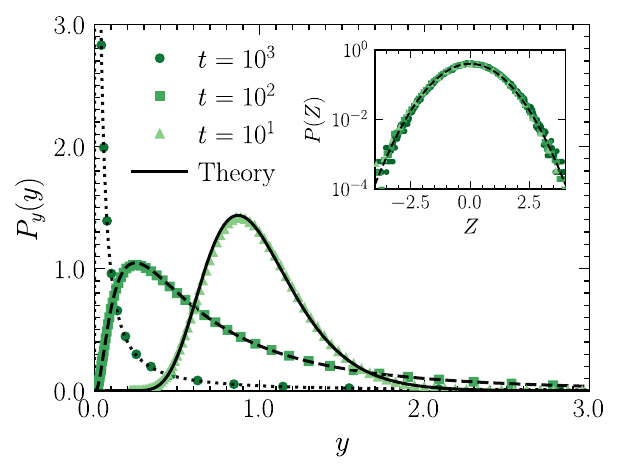}
    \caption{Probability density of the rescaled wealth \(y = W/\langle W\rangle\) at different times $t$, indicated in the legend. Symbols correspond to simulation results at each time $t$.   
    Lines  correspond to the theoretical predictions derived from Eq.~(\ref{eq:lognormal_W}). Inset: data collapse in terms of the standardized log-wealth variable $Z$. Parameter values are $A =  0.01$ and  $D \simeq 4.53 \times 10^{-3}$.
    }
    \label{fig:pdf-rescaled}
\end{figure}

Transforming back via the change of variable $W=e^X$, hence $P_W(W,t) = W^{-1} P_X(\ln W, t)$, we obtain the time-dependent wealth probability density function 
\begin{equation}
P_W(W,t)=\frac{1}{W\sqrt{4\pi Dt}}
\exp\!\left[-\frac{\bigl(\ln W-\ln W_0-C\,t\bigr)^2}{4Dt}\right].
\label{eq:lognormal_W}
\end{equation}
Equation~\eqref{eq:lognormal_W} therefore corresponds to a lognormal distribution in $W$, with a variance that grows diffusively and a mean that drifts linearly in time, namely,
\begin{align}
    \mu(t) & \equiv \ln W_0 + C \,t,\\
    \sigma^2(t) & \equiv 2D\,t,
\end{align}
showing that, in the absence of redistribution, the PDF keeps spreading and does not approach a stationary limit.

Figure~\ref{fig:pdf-rescaled} presents the 
PDF $P(y,t)$ of the mean-normalized wealth  $y=W/\langle W \rangle$, for three different  times. 
showing excellent agreement between agent-based simulations and the corresponding theoretical curves, which follow from  Eq.~(\ref{eq:lognormal_W}). 
Notice that, as time passes, the population undergoes a progressive condensation into the poorest wealth layer.
The inset present the collapse of the solutions using the scaled variable
\begin{equation}
Z(t)=\frac{\ln W(t)-\mu(t)}{\sigma(t)},
\end{equation}
exhibiting the Gaussianity   predicted by 
Eq.~(\ref{eq:fp_X_no_redist_p0}) for $\ln W$.  

\section{Dynamics with redistribution}

When a nontrivial redistribution scheme is present, the discrete-time dynamics is governed by Eq.~\eqref{eq:mapping_p0}. For analytical convenience, we assume that the redistribution step returns the entire tax revenue collected at each time step. Accordingly, we define the redistribution weights as
\begin{equation}
r_{i,t}
=\frac{\phi(y_{i,t})}{\sum_{j=1}^N \phi(y_{j,t})},
\label{eq:redistribution_def}
\end{equation}
which guarantees \(\sum_i r_{i,t}=1\). We also introduce the mean-normalized wealth
\begin{equation}
y_{i,t}=\frac{W_{i,t}}{\langle W \rangle_t},
\label{eq:def_y}
\end{equation}
where \(\langle W \rangle_t\equiv \frac{1}{N}\sum_{j=1}^N W_{j,t}\) denotes the population mean wealth at time \(t\). Neatly, \(\frac{1}{N}\sum_i y_{i,t}=1\) by construction.

To decouple the $N$ maps in Eq.~\eqref{eq:mapping_p0}, we adopt a mean-field closure in which the fluctuating tax pool is replaced by its typical value, namely, 
\begin{equation}
\sum_{j=1}^N f_{j,t}W_{j,t}\approx N \nu\,\langle W \rangle_t.
\label{eq:mf_tax_pool}
\end{equation}
This approximation becomes accurate in the large-\(N\) limit when correlations are weak.
Then, substituting Eq.~\eqref{eq:mf_tax_pool} into Eq.~\eqref{eq:mapping_p0} yields 
\begin{equation}
W_{t+1}
= (1-A)\,\bigl(\nu+\varepsilon\eta_t\bigr)\,W_t
+ A \nu N\,r_t\,\langle W \rangle_t. 
\label{eq:decoupled_mapping_W}
\end{equation}
Using the approximation \(\langle W \rangle_{t+1}\approx \nu\,\langle W \rangle_t\), together with  Eq.~\eqref{eq:redistribution_def}, we obtain the approximate one-agent mapping for the rescaled wealth 
\begin{equation}
y_{t+1}\approx y_t - A\left[y_t-\frac{\phi(y_t)}{\langle \phi \rangle_{t}}\right]
+ s \, y_t\,\eta_t,
\label{eq:y_mapping}
\end{equation}
where 
\begin{equation}
\langle \phi \rangle_{t}\equiv \frac{1}{N}\sum_{j=1}^N\phi(y_{j,t}),
\end{equation}
and
\begin{equation}
s\equiv s(D,A) \equiv \varepsilon\frac{1-A}{\nu} 
\equiv \sqrt{2D}(1-A).
\end{equation}

Taking the continuous-time limit of Eq.~\eqref{eq:y_mapping} in the It\^o sense leads to the SDE
\begin{equation}
\mathrm{d}y
= A\left[\frac{\phi(y)}{\langle\phi\rangle_t}-y\right]\mathrm{d}t
+ s\, y\,\mathrm{d}B_t,
\label{eq:SDE_general}
\end{equation}
where \(B_t\) is a standard Wiener process and
\(\langle\phi\rangle_t \equiv \langle \phi(y(t))\rangle\) denotes the ensemble average of \(\phi(y)\) at time \(t\).
The associated Fokker--Planck equation (FPE) for the PDF \(P_y(y,t)\) reads
\begin{equation}
\partial_t P_y
=A \,\partial_y\!\left[\left(y-\frac{\phi(y)}{\langle\phi\rangle_t}\right)P_y\right]
+ \frac{s^2}{2}\,\partial_y^2\!\bigl(y^2 P_y\bigr).
\label{eq:fp_p0_nonuniform}
\end{equation}

In contrast with the case without redistribution, in the long-time limit, a stationary state is attained (Appendix \ref{app:existence}), characterized by a time-independent density \(P_{y, \rm st}\) and a vanishing probability current \(\partial_t P_y = 0\). Defining 
\begin{equation} \label{eq:alfa}
 \alpha \equiv 2A/s^2 \equiv \frac{A}{D(1-A)^2}, 
\end{equation}
the stationary solution can be written as

\begin{equation}
P_{y,{\rm st}}(y)
=K\,y^{-2-\alpha}
\exp\!\left[\frac{\alpha}{\langle\phi\rangle_{\rm st}}
\int \frac{\phi(y)}{y^2}\,\mathrm{d}y\right],
\label{eq:pst_general}
\end{equation}
where \(K\) is a normalization constant (when it exists) and
\begin{equation}
\langle\phi\rangle_{\rm st}
=\int_0^\infty \phi(y)\,P_{y,{\rm st}}(y)\,\mathrm{d}y
\label{eq:stationary_redistribution}
\end{equation}
must be determined self-consistently. 
Equations~\eqref{eq:pst_general}-\eqref{eq:stationary_redistribution} thus define a stationary law that depends parametrically on the choice of \(\phi\).
Moreover, as shown in Appendix~\ref{app:existence}, for any redistribution kernel $\phi(y)$ that decreases with $y$, the asymptotic tail behavior is given by $P_{y,st}(y) \sim 1/y^{2+\alpha}$.

It can be also useful to describe the approach to stationarity in terms of moments \(m_n(t)=\langle y^n\rangle\). Applying It\^o's lemma \cite{Gardiner2009} to \(y^n\) under Eq.~\eqref{eq:SDE_general} yields, 
\begin{equation}
\frac{\mathrm{d} m_n}{\mathrm{d}t}
=nA\left[\frac{ (n-1)}{\alpha} -1\right]m_n
+ nA\,\frac{\langle y^{\,n-1}\phi(y)\rangle_t}{\langle\phi\rangle_t},
\label{eq:moments_general}
\end{equation}
with \(m_0\equiv 1\).
In particular, for \(n=2\), using \(\langle y(t)\rangle=m_1(t)=1\) for all \(t\), one obtains
\begin{equation}
\frac{\mathrm{d} m_2}{\mathrm{d}t}
=2A\frac{\alpha-1}{\alpha}\,m_2
+ 2A\,\frac{\langle y\,\phi(y)\rangle_t}{\langle\phi\rangle_t},
\label{eq:second_moment}
\end{equation}
which is exact but, in general, not closed, since it involves the nonlinear functional \(\langle y\,\phi(y)\rangle_t\).

The effect of the redistribution mechanism on wealth inequality can be quantified by the Gini coefficient.
Its value ranges from \(0\), corresponding to perfect equality, to \(1\), which represents an extreme oligarchic situation in which all wealth is concentrated in a single agent.
For a population of \(N\) agents with ordered mean-normalized wealth
\(y_1\le y_2\le \cdots \le y_N\), the Gini coefficient is computed numerically as~\cite{Damgaard2000}
\begin{equation}
G
=
\frac{1}{N}
\left[
N+1
-
2\,\frac{\sum_{i=1}^N (N+1-i)y_i}{\sum_{i=1}^N y_i}
\right].
\label{eq:gini_discrete}
\end{equation}
Since the variables are mean-normalized, \(\sum_{i=1}^N y_i=N\).
An equivalent continuum formulation is given by~\cite{Boghosian2015}
\begin{equation}
G(t)
=
1
-
2\int_0^\infty y\,P(y,t)\,S(y,t)\,dy,
\label{eq:gini_int}
\end{equation}
where
\begin{equation}
S(y,t)
=
\int_y^\infty P(u,t)\,du
\label{eq:survival_int}
\end{equation}
is the  complementary cumulative distribution function.
The discrete expression \eqref{eq:gini_discrete} is used for simulation data, whereas Eqs.~\eqref{eq:gini_int} and \eqref{eq:survival_int} are more convenient for analytical calculations.
In general, since \(P(y,t)\) is determined by the Fokker--Planck equation~\eqref{eq:fp_p0_nonuniform}, the integrals in Eqs.~\eqref{eq:gini_int} and \eqref{eq:survival_int} cannot be evaluated in closed form.

In the stationary regime, however, the calculation becomes more tractable.
Once the stationary density of the mean-normalized wealth, \(P_{y,\mathrm{st}}(y)\), is known from Eq.~\eqref{eq:pst_general}, the stationary Gini coefficient can be written as
\begin{equation}
G_{\mathrm{st}}
=
1
-
2\int_0^\infty
y\,P_{y,\mathrm{st}}(y)\,S_{\mathrm{st}}(y)\,dy,
\label{eq:gini_stationary}
\end{equation}
and can be evaluated analytically for specific redistribution kernels \(\phi(y)\), or numerically with high accuracy in the general case.

Having established a general mean-field description in terms of an arbitrary redistribution profile \(\phi(y)\), we now turn to two representative and analytically tractable policies.
Firstly, we consider a uniform redistribution scheme, in which all agents receive the same share of the collected tax, independently of their wealth. Secondly, we study  wealth-dependent schemes that grant  a relative advantage to poorer agents in the redistribution dynamics.

\subsection{Uniform redistribution}
\label{sec:flat_redistribution}

Let us consider the case in which each agent receives a fraction \(1/N\) of the total tax revenue, so that \(\phi(y)\equiv 1\). Equation~\eqref{eq:SDE_general} makes the mechanism behind the redistributive dynamics particularly transparent. The SDE reduces to
\begin{equation} \label{eq:sde-uniform}
    \mathrm{d}y
    = A\,(1-y)\,\mathrm{d}t
    + s y\,\mathrm{d}B_t,
\end{equation}
so that the parameter \(A\) plays the role of an elastic (spring) constant, 
giving rise to a linear restoring drift toward the mean \(y=1\), counteracting the multiplicative noise term~\cite{Anteneodo2005}. 
In fact, Eq.~(\ref{eq:sde-uniform}) belongs to the class of multiplicative mean-reverting processes previously studied in the context of finance~\cite{Anteneodo2005}. 

The stationary distribution \eqref{eq:pst_general} reduces to an inverse-gamma law with shape parameter \(\alpha+1\) and scale parameter \(\alpha\),
\begin{equation}
P_{y,{\rm st}}(y)
= K\,y^{-2-\alpha}
e^{-\frac{\alpha}{y}},
\label{eq:pst_uniform}
\end{equation}
with normalization constant
\begin{equation}
K = \frac{\alpha^{\alpha+1}}{\Gamma(\alpha+1)},
\end{equation}
where \(\Gamma(\cdot)\) denotes the Gamma function~\cite{Butkov1968}.  Note that $P_{y,\mathrm{st}}(y)$ depends on a single parameter, $\alpha$.
The distribution becomes increasingly concentrated around $y=1$ as $\alpha$ increases, which occurs for increasing $A$ or decreasing $D$. Physically, larger values of $\alpha$ correspond to a progressively stronger influence of the mean-reversion process relative to the multiplicative fluctuations. 

Further insight into the statistical properties of the distribution can be obtained from its moments. 
The moment equations form a closed hierarchy, and
Eq.~\eqref{eq:moments_general} reduces to
\begin{equation}
\frac{\mathrm{d} m_n}{\mathrm{d}t}
= nA\left[\frac{(n-1)}{\alpha} - 1\right] m_n
+ nA\, m_{n-1},
\label{eq:moments_uniform}
\end{equation}
with \(m_0 \equiv 1\).
In particular, for the second moment, using the initial condition \(m_2(0)=1\), the solution reads
\begin{equation}
m_2(t)
= m_2^{\mathrm{st}} + \bigl[1 - m_2^{\mathrm{st}}\bigr]\,e^{-t/\tau},
\label{eq:snd_moment_time_evolution}
\end{equation}
where
\begin{equation}
m_{2}^{\rm st}=\frac{\alpha}{\alpha -1},
\qquad
\tau=\frac{\alpha}{2A(\alpha-1)}.
\label{eq:m2_stationary_tau}
\end{equation}
Accordingly, the variance \(\mathrm{Var}[y(t)]=m_2(t)-1\) relaxes as
\begin{equation} 
\mathrm{Var}[y(t)]
=\frac{1}{\alpha-1}\Bigl[1-e^{-t/\tau}\Bigr].
\label{eq:variance}
\end{equation}
A finite stationary variance exists only for  $\alpha>1$.
When  $\alpha \le 1$, the stationary density may still be normalizable for \(A>0\), but its second moment diverges, so the system admits no stationary regime in the strict sense of finite variance.
As $\alpha\to1$ from above, the relaxation time \(\tau\) diverges and the approach to stationarity becomes arbitrarily slow, in line with the impending loss of a finite second moment.

\begin{figure}[b!]
\centering
\includegraphics[width=\columnwidth]{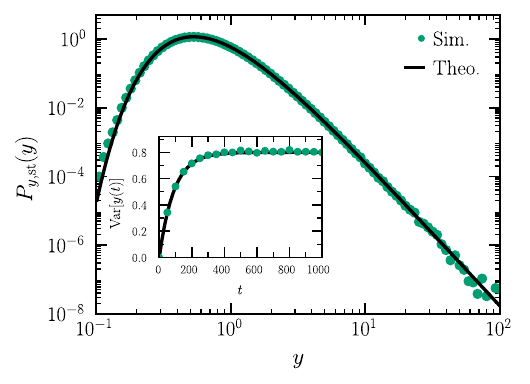}
\caption{\label{fig:pdf-uniform}
Stationary PDF \(P_y(y)\) of the mean-normalized wealth \(y\) for the  uniform redistribution scheme.
Symbols denote results from simulations, while the solid line   shows the corresponding theoretical prediction given by Eq.~\eqref{eq:pst_uniform}.
Parameter values are $A =  0.01$ and  $D \simeq 4.53 \times 10^{-3}$, hence $\alpha = 2.25$. Then, the tail decays approximately as a power law with exponent $4.25$. Inset: time evolution of the variance \(\mathrm{Var}[y(t)]\) compared with the theoretical relaxation given by Eq.~\eqref{eq:variance}. 
}
\end{figure}

In Fig.~\ref{fig:pdf-uniform}, we compare the theoretical stationary distribution  with the results obtained from simulations. The dynamics is evolved up to \(t=10^4\), well beyond the relaxation scale \(\tau  \approx 89.93\), for  \(A = 0.01\) and $D\simeq 4.53 \times 10^{-3}$, which yield  \(\alpha \approx 2.25\). The agreement between theory and simulations is excellent over the full range of \(y\).

The transient dynamics is also consistent with the analytical predictions.
The inset in Fig.~\ref{fig:pdf-uniform} shows that the distribution becomes essentially stationary after a few hundred time steps, and that the variance evolves according to Eq.~\eqref{eq:variance}, 
approaching the stationary value  $1/(\alpha-1) \approx 0.80$, 
in excellent agreement with the simulation results.
Taken together, the simulation results reproduce both the inverse-gamma stationary form \eqref{eq:pst_uniform} and the predicted relaxation of \(m_2(t)\), providing robust evidence in favor of the Fokker--Planck description under uniform redistribution.

Inserting the inverse-gamma stationary density \(P_{y,\mathrm{st}}(y)\) into Eq.~\eqref{eq:survival_int} and performing the change of variables \(v=\alpha/u\), one obtains
\begin{equation}
    S_{\mathrm{st}}(y)
    = \frac{\Gamma(\alpha+1,\alpha/y)}{\Gamma(\alpha+1)},
    \label{eq:survival_uniform}
\end{equation}
where \(\Gamma(v,x)\) denotes the upper incomplete Gamma function.
Substitution into Eq.~\eqref{eq:gini_stationary} then gives
\begin{equation}
    G_{\mathrm{st}}
    =
    1
    - \frac{2\alpha}{\Gamma(\alpha+1)^2}
    \int_0^\infty
   v^{\alpha-1} e^{-v}
    \Gamma(\alpha+1,v)\, dv .
    \label{eq:gini_gamma_integral}
\end{equation}

The remaining integral can be evaluated by writing the incomplete Gamma function in its integral form, exchanging the order of integration, and introducing the transformation $w = v+u$ and $z = u/(v+u)$. 
This change of variables factorizes the integral into a product of a Gamma integral and a Beta integral.
After straightforward algebra, the stationary Gini coefficient for uniform redistribution is obtained in closed form as
\begin{equation}
    G_{\mathrm{st}}
    =
    1 - 2\, I_{1/2}(\alpha+1,\alpha),
    \label{eq:gini_uniform_closed}
\end{equation}
where
\begin{equation}
    I_x(a,b)
    =
    \frac{B_x(a,b)}{B(a,b)}
\end{equation}
is the regularized incomplete Beta function.

\begin{figure}[h!]
\centering
\includegraphics[width=\columnwidth]{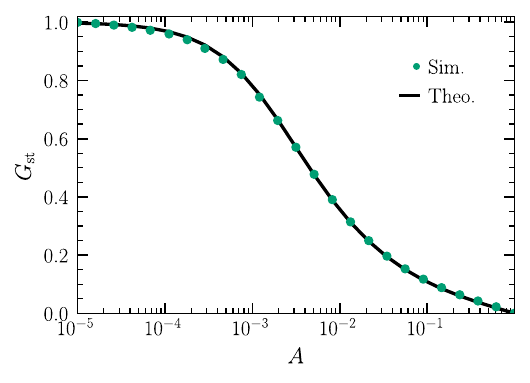}
\caption{\label{fig:gini_vs_A_uniform} Stationary Gini coefficient for uniform redistribution as a function of \(A\).
Symbols denote simulations, while the solid line is the analytical prediction given by Eq.~\eqref{eq:gini_uniform_closed}.
The decrease of \(G_{\mathrm{st}}\) with \(A\) reflects the increasing strength of the redistributive restoring drift relative to multiplicative noise, 
with $D \simeq 4.53 \times 10^{-3}$.
}
\end{figure}

To validate Eq.~\eqref{eq:gini_uniform_closed}, we compare its prediction with results from simulations of the discrete dynamics with uniform redistribution. Figure~\ref{fig:gini_vs_A_uniform} shows the stationary Gini coefficient as a function of the parameter \(A\). 
The analytical curve, obtained from Eq.~\eqref{eq:gini_uniform_closed}, reproduces the simulation results over the whole range of \(A\), confirming that the inverse-gamma stationary solution captures not only the stationary wealth distribution but also the associated inequality level.

Note that \(G_{\mathrm{st}}\) decreases monotonically as \(A\) increases,  
 as expected since larger values of $A$ strengthen the mean-reversion mechanism for fixed amplitude $D$ of the multiplicative fluctuations. 
This reduction in inequality is consistent with the increasing concentration of the probability density function around $y=1$ as $A$ increases, leading to larger values of the exponent $\alpha$.

In the following section, we investigate nonuniform redistribution kernels that preferentially allocate resources toward the poorest fraction of the population.

\subsection{Continuous nonuniform redistribution}
\label{sec:nonuniform_redistribution}

Let us consider a  redistribution kernel that is a smooth monotonically decreasing function of $y$, meaning that the allocation of money is higher for poorer agents, namely,

\begin{equation}
\phi(y)=1 + \frac{\beta}{1+y/y_c},
\label{eq:phi_nonlinear}
\end{equation}
with  \(\beta\ge 0\), \(y_c >0\), so that the redistribution weight is a nonnegative decreasing function  for all \(y\in[0,\infty)\)~\footnote{From a purely analytical standpoint, one could allow \(\phi(y)<0\) without encountering any formal obstruction. However, since taxation and redistribution are modeled as distinct mechanisms, we impose the non-negativity of \(\phi\) to preserve its interpretation as a \textit{bona fide} redistribution weight.}.
Notice that this kernel has two components: a constant term and a nonuniform one. As $\beta$ increases, the nonuniform component becomes increasingly dominant.
Therefore, in the opposite case, \(\beta=0\),  as well as in the limits  $y_c \to 0$, $y_c \to \infty$,  the kernel reduces to the constant profile, corresponding to 
the uniform redistribution   studied in Section~\ref{sec:flat_redistribution}.

The stationary distribution obtained from Eq.~\eqref{eq:pst_general} for this redistribution kernel takes the form
\begin{equation}
P_{y,{\rm st}}(y)
= K\,y^{-2-\alpha-\delta}\bigl(1+y/y_c\bigr)^{\delta}\,
e^{- \frac{\lambda \alpha}{y}},
\label{eq:pst_nonlinear}
\end{equation}
where
\begin{equation}
\lambda =\frac{1+\beta}{\langle \phi\rangle_{\mathrm{st}}}, \;\;\;\; \delta \equiv \frac{\alpha\beta}{\langle \phi\rangle_{\mathrm{st}}}\,\frac{1}{y_c}.
\end{equation}
If $\beta=0$, then $\delta=0$, otherwise $\delta$ is determined self-consistently by the transcendental equation
\begin{equation}
\frac{\alpha\, \beta}{y_c\, \delta}
=\langle \phi\rangle_{\mathrm{st}}= 
1
+
\beta \, (1+\alpha)
\frac{
U\!\left(\alpha+2,\alpha+\delta+2,
\delta \frac{1+\beta}{\beta}\right)
}{
U\!\left(\alpha+1,\alpha+\delta+2,
\delta\,\frac{1+\beta}{\beta}\right) 
},
\label{eq:beta_selfconsistency}
\end{equation}
where \(U(a,b,z)\) denotes the Tricomi confluent hypergeometric function~\cite{abramowitz1964handbook}, and can be computed numerically.
Finally, the normalization constant \(K\) is given by
\begin{equation}
K
= \frac{y_c^{1+\alpha+\delta}}{\Gamma(\alpha+1)\,
U\!\bigl(\alpha+1,\alpha+\delta+2, \delta\,\frac{1+\beta}{\beta}\bigr)}.
\label{eq:normalization_nonlinear}
\end{equation}

 \begin{figure}[t!]
\centering
\includegraphics[width=\columnwidth]{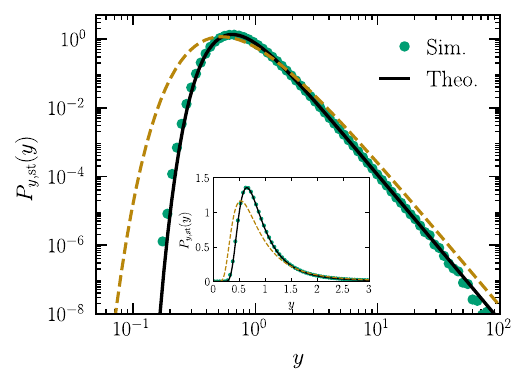}
\caption{\label{fig:stationary_p_0_r_nonlinear}
Stationary PDF of the mean-normalized wealth \(y\) for the nonuniform redistribution scheme given by Eq.~\eqref{eq:phi_nonlinear}.
Symbols denote  the result of simulations at $t=10^4$ steps, the solid line is the theoretical prediction given by Eq.~\eqref{eq:pst_nonlinear} with normalization in Eq.~\eqref{eq:normalization_nonlinear}. 
For comparison, the light dashed line shows the theoretical prediction for the uniform redistribution case, given by Eq.~\eqref{eq:pst_uniform}. 
The inset shows the same plots in linear scales. The parameter values are
\(A=0.01\),  $D \simeq 4.53 \times 10^{-3}$ 
(hence \(\alpha\simeq 2.25\)), \(\beta=10^3\) 
and \(y_c=0.1\).
}
\end{figure}

Numerical simulations corroborate the theoretical predictions, as illustrated in Fig.~\ref{fig:stationary_p_0_r_nonlinear}, which compares the stationary distribution of the mean-normalized wealth $y$ obtained from simulations of the nonuniform redistribution scheme, sampled at $t=10^4$  steps, with the theoretical prediction given by Eqs.~\eqref{eq:pst_nonlinear}--\eqref{eq:normalization_nonlinear}. 
 The agreement is excellent over the entire range of the distribution.

As shown in Appendix~\ref{app:existence}, the stationary distribution exhibits the asymptotic behavior
$P_{y,{\rm st}}(y) \sim y^{-2-\alpha}$ for $y \gg y_c$.
In the opposite limit, \(y \ll y_c\), one has
\begin{equation}
P_{y,{\rm st}}\sim y^{-2-\alpha-\delta}e^{- \frac{\lambda \alpha}{y}},
\qquad y\to 0^+,
\end{equation}
and the inverse-exponential factor provides an essential cutoff that suppresses the probability of extremely low wealth, ensuring integrability at the origin. 
Thus, nonuniform redistribution reshapes the probability density function relative to the uniform case. Although the tail exponent remains determined solely by $\alpha$ as in the uniform case, the  prefactor can be reduced, lowering the probability of finding the wealthiest individuals.  
At the same time, the suppression of low-$y$ values becomes more pronounced than in the uniform case as the coefficient of $1/y$ in the argument of the exponential is larger
(because $\lambda\ge 1$, for any $y_c$ and $\beta$), leading to a stronger cutoff at small wealth.
As a consequence, the distribution tends to become more concentrated than in the homogeneous case, as can be clearly seen in the example presented in Fig.~\ref{fig:stationary_p_0_r_nonlinear}.

\begin{figure}[t!]
\centering
\includegraphics[width=\columnwidth]{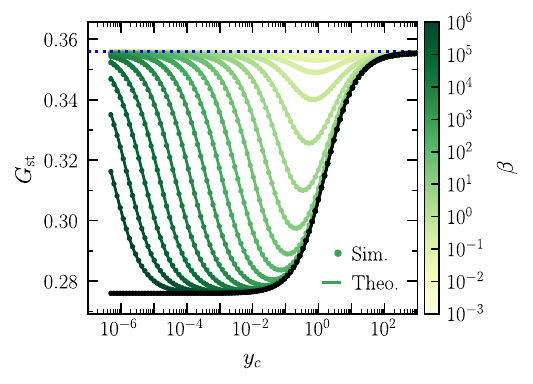}
\caption{
{\bf Continuous nonuniform redistribution.} 
Stationary Gini coefficient as a function of the redistribution scale \(y_c\) for different values of the policy parameter \(\beta\), 
with  \(A=0.01\). 
Symbols denote  simulations, while the solid lines represent the corresponding theoretical prediction from 
Eq.~(\ref{eq:pst_nonlinear}) substituted into Eq.~(\ref{eq:gini_stationary}), in excellent agreement. 
The black line and symbols correspond to the limiting case $\beta\to\infty$ (i.e., $\phi(y)=1/[1+y/y_c]$). 
The horizontal dotted blue line indicates the value of the Gini coefficient for the uniform redistribution kernel. 
Simulations were performed for $N=10^4$ agents,  and recorded at \(t=10^4\)  steps. 
}
\label{fig:gini-nonuniform}
\end{figure}

When increasing $A$, and consequently the amount collected through taxation, we observe the same qualitative picture displayed in Fig.~\ref{fig:gini_vs_A_uniform} for the uniform case, 
but with a lower overall level of the stationary Gini coefficient (not shown). 

 The effect of the kernel parameters ($\beta$ and $y_c$) on inequality is shown in Fig.~\ref{fig:gini-nonuniform}, which  exhibits  the stationary Gini coefficient,  obtained by substituting Eq.~(\ref{eq:pst_nonlinear})  into Eq.~(\ref{eq:gini_stationary}), as a function of the redistribution parameters. 
First, note that the uniform case (recovered for vanishing $\beta$ or vanishing $y_c$ or infinite $y_c$) provides an upper bound for inequality,  \(G_{\mathrm{st}}\simeq 0.3557\), the analytical result derived for uniform redistribution.   

For nonzero $\beta$, a clear non-monotonic dependence on the redistribution scale $y_c$ emerges, consistent with the fact that the uniform-case limit is recovered for extreme values of $y_c$. The Gini coefficient exhibits a minimum at a value of $y_c$ that shifts toward smaller values as the relative weight of the uniform component decreases by increasing $\beta$. At the same time, the minimum becomes progressively broader and shallower.

\subsection{Two-level redistribution}
\label{sec:2level_redistribution}

Let us consider the two-level kernel
\begin{equation} \label{eq:twolevel}
 \phi(y) =
\begin{cases}
1+\beta, & y\le y_c,\\
1, & y>y_c,
\end{cases}
\end{equation}
where \(y_c\) denotes the characteristic wealth scale that controls the onset of selective redistribution. This profile represents an idealized policy with a sharp eligibility threshold: agents below \(y_c\) receive an enhanced redistribution weight, while those above \(y_c\) receive only the baseline share.

For this kernel, Eq.~\eqref{eq:pst_general}, alongside the continuity imposition on the stationary distribution, provides 
\begin{equation}
P_{y,\mathrm{st}}(y)
=
K\,y^{-2-\alpha}
\begin{cases}
\exp\!\left[-\dfrac{\alpha}{\langle \phi\rangle_{\mathrm{st}}}\,
\dfrac{1+\beta}{y}\right], & y\le y_c,\\[4mm]
\exp\!\left[-\dfrac{\alpha}{\langle \phi\rangle_{\mathrm{st}}}
\left(\dfrac{1}{y}+\dfrac{\beta}{y_c}\right)\right], & y> y_c,
\end{cases}
\label{eq:pst_2lvl}
\end{equation}
where the self-consistency condition \eqref{eq:stationary_redistribution} becomes
\begin{equation}
\langle \phi\rangle_{\mathrm{st}}
=
1+\beta \int_{0}^{y_c} P_{y,\mathrm{st}}(y) \, dy.
\label{eq:selfconsistence_2lvl}
\end{equation}
which can be solved numerically by standard root-finding methods.

\begin{figure}[b!]
\includegraphics[width=\columnwidth]{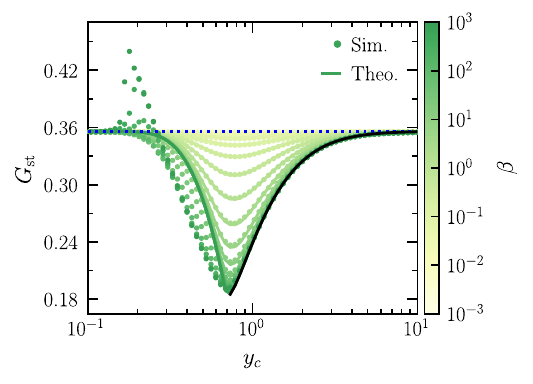} 
\caption{
{\bf Two level redistribution.} Stationary Gini coefficient as a function of the redistribution scale \(y_c\) for different values of the policy parameter \(\beta\). Symbols denote simulations, while the lines represent the theoretical prediction from  Eq.~(\ref{eq:pst_2lvl}) substituted into Eq.~(\ref{eq:gini_stationary}). 
The black line correspond to the limiting case $\beta\to\infty$ (CCT). 
The horizontal dotted blue line indicates the value of the Gini coefficient for the uniform redistribution kernel. 
Simulations were performed for  \(A=0.01\), \(N=10^4\) agents, and recorded at \(t=10^4\) steps.  
}
    \label{fig:twolevel_GINI}
\end{figure}

Figure~\ref{fig:twolevel_GINI} shows the stationary Gini coefficient for the two-level redistribution dynamics as the parameters \(\beta\) and \(y_c\) are varied. Again, we find overall good agreement between theory and  simulations.
A small but systematic deviation, however, becomes visible for large values of \(\beta\) and small \(y_c\), in contrast with the previous cases.
This discrepancy can be understood from the nature of the redistribution dynamics in this region of parameter space: the two-level mechanism generates strong transfer events concentrated on a very small fraction of the population.
Such rare but intense redistribution shocks are not fully captured by the continuous-time approximation underlying the derivation of Eq.~\eqref{eq:SDE_general}, and therefore naturally lead to quantitative deviations between theory and simulation.

The role of \(\beta\) is to control the deviation from the uniform redistribution kernel.
As \(\beta\to0\), the model smoothly recovers the uniform benchmark.
In the opposite limit  \(\beta\to\infty\), the entire redistributed amount is concentrated in the fraction below \(y_c\). Under this limit, the Gini index  reaches its minimum value of \(G_{\mathrm{st}}\simeq 0.18\) at \(y_c \simeq 0.692\).  At this intermediate threshold, a substantial fraction of the population benefits from redistribution, while the upper tail is still sufficiently taxed to transfer wealth from the richest agents toward the bulk of the distribution. This minimum represents the lowest value obtained among the redistribution kernels analyzed in this work for the same values of $A$ and $D$.
It corresponds to an improvement of approximately $49.4\%$ relative to the uniform benchmark and $35.7\%$ relative to the best-performing continuous nonlinear profile, highlighting the effectiveness of a more selective redistribution program.

From a policy perspective, this type of structure, in which redistribution is sharply targeted---so that only agents below the threshold \(y_c\) receive transfers, while those above it are excluded---provides a stylized representation of real-world conditional cash transfer (CCT) and guaranteed minimum income programs.
In particular, this idealized rule may be viewed as a simplified analogue of the Brazilian \textit{Bolsa Fam\'ilia} (BF) program, which is based on direct income transfers targeted at low-income families and conditioned on compliance with health and education requirements.
 Within the model, both the analytical prediction for the CCT limit (solid black curve in Fig.~\ref{fig:twolevel_GINI}; see Appendix~\ref{app:CCT} for the derivation) and the agent-based simulations show that the impact of targeted transfers is governed not by selectivity alone, but by the combined action of the cutoff scale, the redistribution intensity, and the total amount of collected wealth.

Taken together, these findings show that the two-level redistribution scheme can outperform all other kernels considered in this work, but only when selectivity is tuned appropriately.
Its effect is intrinsically nonmonotonic: moderate thresholds maximize inequality reduction, whereas excessively low or high values of \(y_c\) either overconcentrate transfers or drive the system back toward the uniform limit.

\section{Discussion}
\label{sec:discussion}

In the present work, we investigated a slightly altered version of the P.M.C. de Oliveira model \cite{deOliveira2017} for wealth evolution in the case of proportional taxes ($p=0$, $A\neq0$) without non-taxable minimum, and we generalized the  redistribution protocol. The combination of multiplicative growth and a flat tax, whereby every agent pays the same fraction of their wealth, yields a simple model of wealth dynamics that is amenable to analytical treatment and particularly well suited for studying the effects of redistribution policies on inequality, while ensuring sustained economic growth.

As of 2021, proportional, or flat-tax systems, were applied in about $20$ countries at different tax rates, particularly in Central and Eastern Europe (e.g., Hungary, Bulgaria, Romania, Georgia and Ukraine)~\cite{Tanchev2021},
(among these countries, Bulgaria and  Georgia do not establish a non-taxable minimum income threshold~\cite{Tanchev2021}), but also in countries such as Bolivia and in jurisdictions often regarded as tax havens, including Jersey and Guernsey~\cite{WorldPopulationReview2026}. Moreover, poll taxes constitute an even more restrictive form of proportional taxation, as all individuals pay the same fixed amount regardless of income or wealth. Not surprisingly, the Community Charge (commonly known as the poll tax) introduced by Thatcher's government provoked widespread public opposition and culminated in the Poll Tax Riots of March 31, 1990, a major episode of civil unrest that contributed to Thatcher's subsequent resignation~\cite{Graham2010}.

For this model, we developed a Fokker--Planck theory. Starting from the discrete agent-based dynamics, we derived, under a mean-field closure, an Itô stochastic differential equation for the mean-normalized wealth and the associated Fokker--Planck equation. This framework provides a continuous-time description that connects microscopic parameters (tax rate, noise amplitude, and redistribution rule) to macroscopic features of the wealth distribution, such as the existence of a stationary state and its asymptotic tail behavior.

Consistent with previous simulation results~\cite{Barros2025}, 
we show analytically that log-normal wealth distributions emerge in the absence of redistribution ($r_{i,t}=0$). The unbounded growth of the variance of these distributions over time is a signature of ever-increasing inequality, or ''absolute oligarchy formation'', at the critical point of the de Oliveira model ($p=0$)~\cite{deOliveira2017,deOliveira2020}.
We also found that $\langle \ln W \rangle$, the average logarithmic wealth, grows linearly in time whenever the flat tax rate $A$ remains below a threshold value (approximately $1-1/\nu$), determined by the mean wealth return $\nu$. Thus, sustained wealth growth is ensured provided taxation is not excessively high, in agreement with the results previously reported in Ref.~\cite{Barros2025}. Empirical support for our findings is provided by the work of Tanchev \cite{Tanchev2021}. He showed that Bulgaria's flat-tax system, based on a low income tax rate (10\%) and no non-taxable minimum, contributed to rising inequality over the 2008--2019 period rather than reducing it. At the same time, both net average income and the minimum wage increased.

Regarding redistribution, we have shown that both uniform ($r_{i,t}=1/N$) and nonuniform redistribution mechanisms stabilize wealth dynamics, leading to stationary wealth distributions with finite variances. 
As a consequence, inequality no longer grows without bound, and the phase transition to the absorbing state $\omega_{\rm max}=1$ (complete wealth condensation) is suppressed, in agreement with the findings of Refs.~\cite{deOliveira2017,Barros2025}.

Under uniform redistribution, we found that the stationary wealth distribution is an inverse-gamma law controlled by a single parameter, $\alpha=A/[D(1-A)^2]$, which combines the effects of the restoring force and multiplicative fluctuations.
  
For the class of nonuniform redistribution schemes described by the kernel in Eq.~\eqref{eq:phi_nonlinear}, the strongest reduction in inequality (lowest Gini coefficient) occurs at intermediate values of $y_c$ for any value of $\beta$. The global minimum is attained in the limit $\beta \to \infty$, corresponding to the scheme where the uniform redistribution component becomes negligible. 

 For all values of the kernel parameters, however, the tail of the wealth distribution retains the asymptotic form $1/y^{2+\alpha}$, the same of the uniform case. 
Therefore, the the tail is governed solely by the underlying multiplicative growth and taxation dynamics reflected in $\alpha$.
But this nonuniform redistribution reduces the prefactor of the power-law tail and strengthens the low-wealth cutoff, thereby increasing the concentration of agents around the mode of the distribution and effectively enlarging the middle class, such that, in the steady state, the Gini coefficient is reduced for any values of the kernel parameters.
  
For the two-level kernel, the results remain qualitatively similar to those of smooth nonuniform redistribution as long as $A$ is sufficiently small and $\beta$ is not too large. 
Although highly simplified, the two-level redistribution model in the high-\(\beta\) regime provides a useful benchmark for isolating the effects of selectivity in the redistribution mechanism and study CCT programs. 
However, for a range of threshold values $y_c$, whose extent depends on $A$,  the agent-based simulation presents a distinct regime  in which the stationary Gini coefficient $G_{\rm st}$ exceeds the value obtained for the uniform redistribution kernel, as consequence of the discrete time dynamics. From the theoretical standpoint, the $\beta\to\infty$ limit presents the lowest level of inequality at $y_c \simeq 0.692$, with $G_{\rm st}\simeq 0.18$. 
Therefore, the effectiveness of this kernel is conditioned by the values of the parameters.

Within the framework of the multiplicative market dynamics considered here, we have shown that a broad class of redistribution mechanisms gives rise to stationary wealth distributions that exhibit remarkably similar qualitative stylized facts. This robustness suggests the existence of a certain degree of universality, whereby the large-scale statistical properties of the system are largely insensitive to the specific details of the redistribution rule. 

Such a finding may provide a possible explanation for the apparent universality often observed in empirical wealth and income distributions across countries operating under diverse tax systems and fiscal policies. In this view, the overall shape of the distribution may be determined more by the underlying market dynamics than by the details of the redistribution mechanism.  

An interesting direction for future research would be to investigate whether a similar universality also emerges in other classes of market models, such as those based on yard-sale dynamics. Another important extension would be to generalize the present framework to incorporate nonlinear taxation schemes, allowing for a more general description of redistribution policies and their impact on economic inequality.

\section*{Acknowledgments}

We acknowledge partial financial support from the \textit{Coordena\c{c}\~ao de Aperfei\c{c}oamento de Pessoal de N\'{\i}vel Superior} (CAPES), Brazil, under Finance Code 001 and Grant No. 88887-708770/2022-00. M.L. also acknowledge partial financial support from the \textit{Funda\c{c}\~ao de Amparo \`a Pesquisa do Estado de Minas Gerais} (FAPEMIG), Brazil, under Grant No. APQ-01973-24. C.A. also acknowledges partial financial support from the \textit{Conselho Nacional de Desenvolvimento Cient\'{\i}fico e Tecnol\'ogico} (CNPq), Brazil, under Grants No. 308347/2025-0 and No. 406820/2025-2 (Universal), and from the \textit{Funda\c{c}\~ao de Amparo \`a Pesquisa do Estado do Rio de Janeiro} (FAPERJ), Brazil, under Grant No. CNE E-26/204.130/2024.

\appendix
\section{Existence of stationary regime}
\label{app:existence}

In order to ensure that \(r_{t}\) represents a plausible wealth-based redistribution rule, we imposed three minimal conditions on the profile \(\phi(y)\), besides integrability and continuity:
(i) nonnegativity, so that each agent receives a nonnegative share of the tax pool;
(ii) a positive floor, so that the poorest agents always receive some transfer; and
(iii) non-regressivity, in the sense that richer agents do not receive a larger amount than poorer ones.
Mathematically, we translate these requirements into
\begin{equation}
\phi(0)=\phi_0>0,
\qquad
\frac{d\phi}{dy}\leq 0.
\label{eq:phi_conditions}
\end{equation}
In this appendix we will show that these conditions are sufficient to guarantee the existence of a stationary solution of the Fokker--Planck equation. 

A stationary regime exists provided,
\begin{equation} 
\int_0^\infty y^{-2-\alpha} \exp\!\left[ \frac{\alpha}{c}\int^y \frac{\phi(u)}{u^2}\,du \right] dy < \infty.
\label{eq:normalization_condition} \end{equation}
The convergence of \eqref{eq:normalization_condition}, depends on its behavior near the endpoints \(y\to0\) and \(y\to\infty\). So, we split the integral, 
for instance, as follows
\begin{equation}
\begin{aligned}
I
&= \int_0^1 P_{y, {\rm st}}(y)\,dy
 + \int_1^\infty P_{y, {\rm st}}(y)\,dy \\
&\equiv I_< + I_>,
\end{aligned}
\label{eq:I_split}
\end{equation}
so that convergence of both \(I_<\) and \(I_>\) guarantees the existence of a normalized stationary density.

We begin by examining the integrability of the stationary density in the low-wealth limit.
From continuity and the conditions in Eq.~\eqref{eq:phi_conditions}, it follows that
\(\phi(y)\to\phi_0>0\) as \(y\to0\).
Hence, for sufficiently small \(y\),
\begin{equation}
\int^y \frac{\phi(u)}{u^2}\,du
\simeq -\frac{\phi_0}{y} + \mathrm{const},
\end{equation}
which implies
\begin{equation}
P_{y,\mathrm{st}}(y)
\propto y^{-2-\alpha}
\exp\!\left(-\frac{\alpha\phi_0}{c\,y}\right),
\qquad y\to0.
\label{eq:pst_y_to_0}
\end{equation}
Accordingly, the contribution from the interval \((0,1)\) behaves as
\begin{align}
I_<
&\sim \int_0^1
y^{-2-\alpha}
\exp\!\left(-\frac{\alpha\phi_0}{c\,y}\right)\,dy \nonumber\\
&=
\int_0^1
\exp\!\left[
-\frac{\alpha\phi_0}{c\,y}
-(2+\alpha)\ln y
\right]\,dy.
\end{align}
Since \(1/y\) diverges faster than \(|\ln y|\) as \(y\to0\), the exponential factor
\(\exp[-\alpha\phi_0/(c\,y)]\) dominates the asymptotic behavior and suppresses the integrand superexponentially.
Therefore, \(I_<\) is finite.

We now turn to the large-wealth limit. From Eq.~\eqref{eq:phi_conditions}, the redistribution profile \(\phi(y)\) is nonincreasing and bounded above by \(\phi_0\), so that
\begin{equation}
0 \leq \phi(u) \leq \phi_0
\Rightarrow 
\int_1^\infty \frac{\phi(u)}{u^2}\,du
\leq
\phi_0 \int_1^\infty \frac{du}{u^2}
=
\phi_0.
\end{equation}
It follows that the integral \(\int^y \phi(u)/u^2\,du\) converges to a finite limit as \(y\to\infty\), and therefore the exponential factor  approaches a positive constant. As a result, the stationary density has the asymptotic form
\begin{equation}
P_{y,\mathrm{st}}(y)
\sim C\,y^{-2-\alpha},
\qquad y\to\infty,
\label{eq:pst_y_to_infty}
\end{equation}
for some constant \(C>0\). Since \(\alpha>0\), the tail integral
\begin{equation}
\int_1^\infty y^{-2-\alpha}\,dy
\end{equation}
is convergent, which implies that \(I_>\) is finite.

Combining this result with the analysis of the \(y\to0\) sector, we conclude that both contributions \(I_<\) and \(I_>\) remain finite under the conditions in Eq.~\eqref{eq:phi_conditions}. Therefore, the stationary density \(P_{y,\mathrm{st}}(y)\) is normalizable. Equivalently, any redistribution rule \(\phi(y)\) satisfying the minimal physical requirements of nonnegativity, a positive floor, and non-regressivity leads, within the present Fokker--Planck description, to a well-defined stationary regime with finite normalization.

\section{CCT limit}
\label{app:CCT}

In the limit of an infinitely selective redistribution mechanism, corresponding to
\(\beta\to\infty\), the two-level kernel reduces to the idealized
\textit{conditional cash transfer} (CCT) profile
\begin{equation}
\phi_{\mathrm{BF}}(y)=
\begin{cases}
1, & y\le y_c,\\
0, & y>y_c,
\end{cases}
\label{eq:BF}
\end{equation}
so that only agents with wealth below the threshold \(y_c\) receive redistributed resources.
Substituting Eq.~\eqref{eq:BF} into the general stationary solution, Eq.~\eqref{eq:pst_general}, yields
\begin{equation}
P_{\mathrm{BF}}(y)
=
K\,y^{-2-\alpha}
\begin{cases}
\displaystyle
\exp\!\left[
-\frac{\alpha}{f_{\mathrm{st}}}
\left(
\frac{1}{y}-\frac{1}{y_c}
\right)
\right],
& y\le y_c,\\[1.2ex]
1, & y>y_c,
\end{cases}
\label{eq:pst_bf}
\end{equation}
where
\begin{equation}
f_{\mathrm{st}}
=
\langle \phi\rangle_{\mathrm{st}}
=
\int_0^{y_c} P_{\mathrm{BF}}(y)\,dy
\label{eq:selfconsistency_bf}
\end{equation}
is the stationary fraction of beneficiaries, and \(K\) is fixed by normalization.
The theoretical CCT curve shown in Fig.~\ref{fig:twolevel_GINI} is obtained by solving
Eq.~\eqref{eq:selfconsistency_bf} numerically and then evaluating the Gini coefficient from the resulting stationary density.
As expected, this sharply targeted limit agrees closely with the high-\(\beta\) curves.

The existence of the CCT stationary solution is, however, restricted to \(y_c>y_c^\ast\).
This threshold can be derived analytically by considering the limit \(f_{\mathrm{st}}\to0\).
In Eq.~\eqref{eq:pst_bf}, the exponential factor for \(y<y_c\) attains its maximum at the endpoint \(y=y_c\) and is exponentially suppressed elsewhere.
Hence, the integral is dominated by a narrow boundary layer near \(y_c\).
Applying Laplace's method---equivalently, Watson's lemma for an endpoint maximum---one finds the leading-order contribution
\begin{equation}
\int_0^{y_c}
y^{-2-\alpha}
\exp\!\left[
-\frac{\alpha}{f_{\mathrm{st}}}
\left(
\frac{1}{y}-\frac{1}{y_c}
\right)
\right]
dy
\sim
\frac{f_{\mathrm{st}}}{\alpha}\,y_c^{-\alpha}.
\label{eq:watson_result}
\end{equation}
Combining this asymptotic form with the normalization condition gives
\begin{equation}
f_{\mathrm{st}}
\sim
f_{\mathrm{st}}\,
\frac{1+\alpha}{\alpha}\,y_c.
\label{eq:self_bf_asymptotic}
\end{equation}
Consequently, a nonzero solution can exist only if
\begin{equation}
y_c>y_c^\ast,
\qquad
y_c^\ast
=
\frac{\alpha}{1+\alpha}.
\label{eq:yc_critical_bf}
\end{equation}

For \(y_c\le y_c^\ast\), the self-consistency equation no longer admits a
physically acceptable solution with \(f_{\mathrm{st}}>0\) and \(\langle y\rangle_{\mathrm{st}}=1\).
In this regime, redistribution is concentrated on a vanishing fraction of agents,
and the transfer received by each beneficiary diverges as \(1/f_{\mathrm{st}}\). Thus, the continuous (\(y\ge y_c\)) component carries only a fraction 
\begin{equation}
\langle y\rangle
=
\frac{1+\alpha}{\alpha}\,y_c
=
\frac{y_c}{y_c^\ast}
<1.
\label{eq:bf_regular_mean}
\end{equation}
of the total normalized wealth, while the remaining fraction,
\begin{equation}
1-\frac{y_c}{y_c^\ast},
\end{equation}
must be supported by a singular component, corresponding to wealth condensation in a vanishing fraction of the population.

\newpage

\bibliographystyle{unsrt}
\bibliography{bib}

@PREAMBLE{
 "\providecommand{\noopsort}[1]{}" 
 # "\providecommand{\singleletter}[1]{#1}%" 
}

@book{abramowitz1964handbook,
  title={Handbook of mathematical functions with formulas, graphs, and mathematical tables},
  author={Abramowitz, Milton and Stegun, Irene A},
  volume={55},
  year={1964},
  publisher={Dover Publications},
  address={New York}
}

@Article{Calvelli2023,
AUTHOR = {Calvelli, Matheus and Curado, Evaldo M. F.},
TITLE = {A Wealth Distribution Agent Model Based on a Few Universal Assumptions},
JOURNAL = {Entropy},
VOLUME = {25},
YEAR = {2023},
NUMBER = {8},
ARTICLE-NUMBER = {1236},
URL = {https://www.mdpi.com/1099-4300/25/8/1236},
PubMedID = {37628266},
ISSN = {1099-4300},
DOI = {10.3390/e25081236}
}

@article{deOliveira2017,
doi = {10.1209/0295-5075/119/40007},
url = {https://doi.org/10.1209/0295-5075/119/40007},
year = {2017},
month = {nov},
publisher = {EDP Sciences, IOP Publishing and Società Italiana di Fisica},
volume = {119},
number = {4},
pages = {40007},
author = {de Oliveira, Paulo Murilo Castro},
title = {Rich or poor: Who should pay higher tax rates?},
journal = {Europhysics Letters}
}

@article{deOliveira2020,
  author       = {de Oliveira, Paulo Murilo Castro},
  title        = {Investment/taxation/redistribution model criticality},
  journal      = {The European Physical Journal B},
  year         = {2020},
  volume       = {93},
  number       = {10},
  pages        = {196},
  doi          = {10.1140/epjb/e2020-10308-x},
  url          = {https://doi.org/10.1140/epjb/e2020-10308-x},
  issn         = {1434-6036}
}

@article{Lima2022,
title = {Nonlinear redistribution of wealth from a stochastic approach},
journal = {Chaos, Solitons \& Fractals},
volume = {163},
pages = {112578},
year = {2022},
issn = {0960-0779},
doi = {https://doi.org/10.1016/j.chaos.2022.112578},
url = {https://www.sciencedirect.com/science/article/pii/S0960077922007688},
author = {Hugo Lima and Allan R. Vieira and Celia Anteneodo}
}

@article{Barros2025,
title = {Effects of taxes, redistribution actions and fiscal evasion on wealth inequality: An agent-based model approach},
journal = {Physica A: Statistical Mechanics and its Applications},
volume = {679},
pages = {130960},
year = {2025},
issn = {0378-4371},
doi = {https://doi.org/10.1016/j.physa.2025.130960},
url = {https://www.sciencedirect.com/science/article/pii/S0378437125006120},
author = {Iago N. Barros and Marcelo L. Martins}
}

@article{Tanchev2021,
  author    = {Tanchev, S.},
  title     = {How the proportional income taxation increases inequality in {Bulgaria} },
  journal   = {Journal of Tax Reform},
  year      = {2021},
  volume    = {7},
  number    = {3},
  pages     = {244--254},
  doi       = {10.15826/jtr.2021.7.3.101},
  url       = {https://doi.org/10.15826/jtr.2021.7.3.101}
}

@misc{WorldPopulationReview2026,
  author       = {{World Population Review}},
  title        = {Current World Population 2026},
  year         = {2026},
  url          = {https://worldpopulationreview.com/},
  note         = {Accessed: 2026-05-13}
}

@article{Graham2010,
  author  = {Graham, Daniel},
  title   = {The battle of Trafalgar Square: the poll tax riots revisited},
  journal = {The Independent},
  year    = {2010},
  month   = mar,
  day     = {21},
  url     = {https://www.independent.co.uk/news/uk/politics/the-battle-of-trafalgar-square-the-poll-tax-riots-revisited-1926873.html},
  note    = {Accessed: 2026-05-13}
}

@article{Boghosian2017,
title = {Oligarchy as a phase transition: The effect of wealth-attained advantage in a Fokker–Planck description of asset exchange},
journal = {Physica A: Statistical Mechanics and its Applications},
volume = {476},
pages = {15-37},
year = {2017},
issn = {0378-4371},
doi = {https://doi.org/10.1016/j.physa.2017.01.071},
url = {https://www.sciencedirect.com/science/article/pii/S037843711730081X},
author = {Bruce M. Boghosian and Adrian Devitt-Lee and Merek Johnson and Jie Li and Jeremy A. Marcq and Hongyan Wang}
}

@article{Bouchaud2000,
title = {Wealth condensation in a simple model of economy},
journal = {Physica A: Statistical Mechanics and its Applications},
volume = {282},
number = {3},
pages = {536-545},
year = {2000},
issn = {0378-4371},
doi = {https://doi.org/10.1016/S0378-4371(00)00205-3},
url = {https://www.sciencedirect.com/science/article/pii/S0378437100002053},
author = {Jean-Philippe Bouchaud and Marc Mézard}
}

@article{Toscani2008,
  title = {Kinetic equations modelling wealth redistribution: A comparison of approaches},
  author = {D\"uring, Bertram and Matthes, Daniel and Toscani, Giuseppe},
  journal = {Phys. Rev. E},
  volume = {78},
  issue = {5},
  pages = {056103},
  numpages = {12},
  year = {2008},
  month = {Nov},
  publisher = {American Physical Society},
  doi = {10.1103/PhysRevE.78.056103},
  url = {https://link.aps.org/doi/10.1103/PhysRevE.78.056103}
}

@article{Iglesias2020,
title = {Wealth distribution models with regulations: Dynamics and equilibria},
journal = {Physica A: Statistical Mechanics and its Applications},
volume = {551},
pages = {124201},
year = {2020},
issn = {0378-4371},
doi = {https://doi.org/10.1016/j.physa.2020.124201},
url = {https://www.sciencedirect.com/science/article/pii/S0378437120300406},
author = {Ben-Hur Francisco Cardoso and Sebastián Gonçalves and José Roberto Iglesias}
}

@article{Polk2021,
  author    = {Polk, Sam L. and Boghosian, Bruce M.},
  title     = {The Nonuniversality of Wealth Distribution Tails Near Wealth Condensation Criticality},
  journal   = {SIAM Journal on Applied Mathematics},
  volume    = {81},
  number    = {4},
  pages     = {1717--1741},
  year      = {2021},
  doi       = {10.1137/19M1306051},
  url       = {https://doi.org/10.1137/19M1306051}
}

@article{Boghosian2015,
  author    = {Boghosian, Bruce M. and Johnson, Merek and Marcq, Jeremy A.},
  title     = {An {H} Theorem for {Boltzmann's} Equation for the Yard-Sale Model of Asset Exchange},
  journal   = {Journal of Statistical Physics},
  volume    = {161},
  number    = {6},
  pages     = {1339--1350},
  year      = {2015},
  doi       = {10.1007/s10955-015-1316-8},
  url       = {https://doi.org/10.1007/s10955-015-1316-8},
  issn      = {1572-9613}
}

@article{Damgaard2000,
author = {Damgaard, Christian and Weiner, Jacob},
title = {DESCRIBING INEQUALITY IN PLANT SIZE OR FECUNDITY},
journal = {Ecology},
volume = {81},
number = {4},
pages = {1139-1142},
doi = {https://doi.org/10.1890/0012-9658(2000)081[1139:DIIPSO]2.0.CO;2},
url = {https://esajournals.onlinelibrary.wiley.com/doi/abs/10.1890/0012-9658%282000%29081%5B1139%3ADIIPSO%5D2.0.CO%3B2},
eprint = {https://esajournals.onlinelibrary.wiley.com/doi/pdf/10.1890/0012-9658%282000%29081%5B1139%3ADIIPSO%5D2.0.CO%3B2},
year = {2000}
}

@article{Anteneodo2005,
  title = {Additive-multiplicative stochastic models of financial mean-reverting processes},
  author = {Anteneodo, Celia and Riera, Rosane },
  journal = {Phys. Rev. E},
  volume = {72},
  issue = {2},
  pages = {026106},
  numpages = {7},
  year = {2005},
  month = {Aug},
  publisher = {American Physical Society},
  doi = {10.1103/PhysRevE.72.026106},
  url = {https://link.aps.org/doi/10.1103/PhysRevE.72.026106}
}

@book{Gardiner2009,
author    = {Gardiner, Crispin W.},
title     = {Stochastic Methods: A Handbook for the Natural and Social Sciences},
edition   = {4},
publisher = {Springer},
address   = {New York},
year      = {2009}
}

@book{Butkov1968,
  title={Mathematical Physics},
  author={Butkov, E.},
  isbn={9780201007275},
  lccn={68011391},
  series={A-W series in advanced physics},
  url={https://books.google.com.br/books?id=aGwNlA_i3pkC},
  year={1968},
  publisher={Addison-Wesley Publishing Company}
}

\end{document}